\documentclass[epsfig,psfig,aps,twocolumn,prl]{revtex4-1}
\usepackage[colorlinks=true,citecolor=red,urlcolor=blue]{hyperref}
\usepackage{amsfonts}
\usepackage{graphicx}
\usepackage{epsfig}
\usepackage{epsf}
\begin{document}
\draft
\title
{
Inhomogeneous superconductivity in organic conductors: role of disorder and magnetic field
}
\author
{
S. Haddad$^1$, S. Charfi-Kaddour$^1$ and J.-P. Pouget$^2$
} 
\address{
$^1$Laboratoire de Physique de la Mati\`ere Condens\'ee, D\'epartement de Physique,
Facult\'e des Sciences de Tunis, Universit\'e Tunis El Manar, Campus universitaire 1060 Tunis, Tunisia\\
$^2$Laboratoire de Physique des Solides, CNRS-Universit\'e Paris-Sud (UMR 8502) 91405 Orsay, France
}
%
%
\begin{abstract}
Several experimental studies have shown the presence of spatially inhomogeneous phase coexistence of superconducting and non superconducting domains in low dimensional organic superconductors. The superconducting properties of these systems are found to be strongly dependent on the amount of disorder introduced in the sample regardless of its origin. The suppression of the superconducting transition temperature $T_c$ shows clear discrepancy with the result expected from the Abrikosov-Gor'kov law giving the behavior of $T_c$ with impurities.
Based on the time dependent Ginzburg-Landau theory, we derive a model to account for the striking feature of $T_c$ in organic superconductors for different types of disorder by considering the segregated texture of the system. We show that the calculated $T_c$ quantitatively agrees with experiments.
We also focus on the role of superconducting fluctuations on the upper critical fields $H_{c2}$ of layered superconductors showing slab structure where superconducting domains are sandwiched by non-superconducting regions. We found that $H_{c2}$ may be strongly enhanced by such fluctuations. 
\end{abstract}
\pacs{
PACS numbers:74.70.Kn, 74.20.De, 74.40.+k
}
\maketitle

\section{Introduction}
Low dimensional organic conductors have attracted sustained interest regarding their rich phase diagram showing almost all known ground states with particularly a superconducting phase next to a magnetic ordering \cite{Revue,RevLebed,RevMuller,DeMelo,RevLang}, which is reminiscent of high-$T_c$ superconductors. This proximity is believed to result in a phase segregation consisting of an inhomogeneous coexistence of electronic states such as the nanoscale charge stripes observed in copper oxide \cite{HTC_stripe}. Mesoscopic scale inhomogeneities have been reported in the organic charge transfer salts (TMTSF)$_2$X \cite{Vuletic,LeePF6,PasquierPF6}, so-called Bechgaard salts, and $\kappa$(BEDT-TTF)$_2$Y \cite{RevMuller,Lefebvre} where TMTSF denotes the donor molecule tetramethyl-tetraselenafulvalene while BEDT-TTF stands for the bis(ethylenedithio)tetrathiafulvalene; X=ClO$_4$, PF$_6$, ReO$_4$... and Y=Cu(NCS)$_2$, Cu[N(CN)$_2$]Z (Z=Br and Cl) are inorganic anions.
These compounds are highly anisotropic: the former show marked quasi-one dimensional (q1D) electronic properties whereas the latter, abbreviated as (ET)$_2$, can be regarded as a layered (two-dimensional 2D) systems. The switching between the several low temperature electronic phases of these compounds can be controlled by applying hydrostatic pressure or by chemical substitution of either the inorganic anion or the donor molecule.

A close investigation of the boundary between the magnetic phase and the superconducting one reveals a mesoscopic phase segregation where the system shows coexistence of superconducting domains separated by magnetic ones \cite{RevLang,Vuletic,LeePF6,PasquierPF6,Lefebvre}. Several experimental studies have reported that the interplay between these two phases is substantially dependent on the purity of the sample.
Likewise, it has been found that this segregated structure can be induced by varying the hydrostatic pressure around the critical value $P_c$ at which the magnetic phase collapses \cite{Vuletic,LeePF6,PasquierPF6,Sushko,Lefebvre}.
Disorder turns out to be a key parameter for the stability of the superconducting state in these compounds. Despite the general consensus on the unconventional nature of the superconducting phase of different organic superconductors, the symmetry of the corresponding order parameter is still under debate \cite{RevLebed}. Studying the effect of disorder on the superconducting phase may bring an answer to the puzzling question of the superconducting symmetry.
In these molecular materials, two types of disorder can be found: (i) Extrinsic disorder resulting from the random imperfections obtained by introducing impurities or radiation damages. (ii) Intrinsic disorder related to the noncentrosymmetric character of the inorganic anions, such as ClO$^-_4$ in (TMTSF)$_2$ClO$_4$\cite{Pouget93}, or to the presence of internal degrees of freedom such as partially disordered ethylene groups adopting several conformations in $\kappa$(ET)$_2$X \cite{RevMuller}.
These structural degrees of freedom, which are strongly dependent on the cooling history of the sample, have drastic effect on the superconducting state. A rapid cooling of the sample induces a strong depression of the superconducting transition temperature $T_c$ and may even destroy superconductivity as in (TMTSF)$_2$ClO$_4$\cite{RevLebed}.\

There have been several experimental studies dealing with the disorder in organic superconductors. In (TMTSF)$_2$X, studies were focused particularly on the cooling rate induced disorder in (TMTSF)$_2$ClO$_4$ \cite{Schwenk,Tomic82,JooPhD} and disorder generated by the anion substitution in (TMTSF)$_2$(ClO$_4$)$_{1-x}$(ReO$_4$)$_{x}$ \cite{Ilakovac,Tomic83,Joo05,Joo04}.
In $\kappa$(ET)$_2$X, disorder was introduced by rapid cooling \cite{Su,Stalcup,Yon_cool_04,Yon_cool_05}, chemical substitution of donor or anion \cite{Sasaki05,Yon04,Yon07,Sasaki08,Sasaki10}, X-ray and proton damage \cite{Sasaki10,Tokumoto2,Analytis}.

The superconducting transition temperature T$_c$ is found to be reduced by increasing the amount of disorder regardless of its origin \cite{Tokumoto,Ito91}. A theoretical analysis \cite{Powell} of the dependence of T$_c$ on disorder in different (ET)$_2$X salts has shown that the suppression of T$_c$ obeys to the Abrikozov-Gor'kov (AG) law \cite{AG} which describes the suppression of T$_c$ in the presence of either magnetic impurities or non magnetic impurities in non-s-wave superconductors. However, for large disorder, clear departure from the AG formula has been reported in several low dimensional organic superconductors \cite{Yon04,Yon_cool_04,Analytis,JooPhD}.
Alternative mechanisms have been suggested to account for this peculiar behavior \cite{Analytis,Hasegawa,EPL}. A more detailed discussion is given below.\

The outcomes of the studies dealing with the effect of disorder on organic superconductors are consistent with the unconventional character of the superconducting state since the latter is suppressed in presence of non-magnetic impurities\cite{RevLebed}. However, these studies do not state clearly the symmetry of the superconducting order parameter. 
Experiments carried out on Bechgaard salts \cite{LeePF6,Shinagawa,Yonezawa} show that the upper critical fields $H_{c2}$ along the most conducting axes $a$ and $b^{\prime}$, $H^a_{c2}$ and $H^b_{c2}$, are clearly times larger than the Pauli paramagnetic limit $H_p$ above which $H^a_{c2}$ and $H^b_{c2}$ continue to increase with no sign to saturation.
This behavior seems to rule out the singlet state and gives compelling indications of the presence of a spin triplet superconducting phase.
However, the question of the superconducting gap symmetry in Bechgaard salts is not settled yet. Despite the general tendency to accept the singlet character of the low field state, the symmetry of the high field phase continues to be a controversial issue \cite{RevLebed,RevMuller,Shinagawa,DeMelo}.

In this paper, we first complete the theoretical study presented in Ref.\cite{EPL} dealing with the role of cooling induced disorder on the inhomogeneous superconducting phase of (TMTSF)$_2$ClO$_4$ where it is found that T$_c$ is strongly suppressed as the mixed character is enhanced. We then present a quantitative comparison with experiments which was lacking in Ref.\cite{EPL}. In the present work, we give a detailed analysis of the experimental data in quasi-1D and 2D organic superconductors to infer the key parameters used in our model. We then derive the behavior of T$_c$ as a function of various types of disorder. To the best of our knowledge our results are the first to account quantitatively for the suppression of the superconducting transition temperature by various sources of disorder in several organic superconductors. There is a general consensus, emerging from recent experiments, on the discrepancy between the experimental data obtained at large disorder rate and the results expected from the AG law \cite{Yon04,Yon_cool_04,Analytis,JooPhD}. Our work will shed light on the origin of this discrepancy.
We also briefly present in this paper, the effect of inhomogeneity on the upper critical fields in the case of layered organic superconductors where superconducting domains are sandwiched between non-superconducting regions as it has been reported in (TMTSF)$_2$PF$_6$ \cite{LeePF6,PasquierPF6}.
In the next section, we summarize the experimental results related to the dependence of T$_c$ on disorder in several low dimensional organic superconductors. After a brief comment on the theoretical results presented in literature, we describe, in Section 3, the method we used to extract, from the experimental data, the theoretical parameters of the model we described in Ref.\cite{EPL}. In section 4, results are discussed in connection with experiments.
It is worth noting that, as done in Refs.\cite{Ullah,Puica}, we do not consider a particular symmetry of the superconducting gap. However, the decrease of T$_c$ as a function of non magneticc disorder, reported in the present paper, supports the scenario of unconventional superconducting order parameter in the different low dimensional organic superconductors discussed in this work. This scenario seems to be the subject of a general consensus emerging from the huge experimental results dealing with these compounds \cite{RevLebed,RevMuller,RevSingleton,Carrington}. Nevertheless, the debate on the gap symmetry is not completely settled.

\section{Some experimental facts}

When cooled sufficiently slowly, the ClO$^-_4$ anions in (TMTSF)$_2$ClO$_4$ order for entropy reasons \cite{RevLang,Pouget00} at a structural transition temperature $T_{AO}\sim24$ K. The nature of the electronic ground state, stabilized below $T_{AO}$, is found to be substantially dependent on the cooling rate. In the relaxed state (R-state), obtained for a cooling rate less than 0.1K/min, the ground state is superconducting below 1.2 K \cite{Revue,RevLang}. However, the quenched state (Q-state) resulting of a cooling at a rate more than 50K/min, is an insulating spin density wave (SDW) \cite{Schwenk,Takahashi82,Tomic82,Walsh}. The most striking effect of the cooling rate is obtained for intermediate cooling rates at which the sample exhibits an inhomogeneous mixture of superconducting domains (in which ClO$_4$ anions are ordered) and SDW regions (where ClO$_4$ anions are disordered)\cite{Schwenk,Pouget,Tomic83,JooPhD,Joo04,Garoche,Tomic82,Schwenk83,Meissner,Mat,Greer,Park}.\
Meissner effect signal \cite{Schwenk83,Meissner} showed that, by increasing the cooling rate, the insulating regions get larger at the expense of the superconducting ones. These results have been corroborated by X-ray study \cite{Pouget} showing a variation of the fraction of ordered ClO$_4$ domains as a function of the cooling rate. The fraction value was found to be identical to that of the superconducting volume deduced from the Meissner  measurement  \cite{Mat}. The X-ray study has also provided the average size of these ordered regions as a function of the cooling rate.\\

This Phase segregation has been also induced by X-ray irradiation \cite{Park} and chemical substitution\cite{Tomic83,Joo05} in (TMTSF)$_2$ClO$_4$ and by hydrostatic pressure in (TMTSF)$_2$PF$_46$ \cite{Vuletic,LeePF6,PasquierPF6}.

It is worth to stress that the possibility of a SDW/superconducting phase coexistence in (TMTSF)$_2$PF$_6$ was first suggested by Greene {\it et al.} \cite{Greene} who also considered a filamentary superconducting domain structure.\\

The signature of such coexistence has been also reported in the organic superconductor (MDT-TS)(AuI$_2$)$_{0.441}$, where MDT-TS denotes methylenediothio-tetraselenafulvalene, showing a temperature-pressure phase diagram reminiscent of those obtained in (TMTSF)$_2$X and $\kappa$(ET)$_2$X salts \cite{Kawamoto09}.\\

Several experimental studies have revealed that the cooling rate has a drastic effect on the low temperature electronic properties of $\kappa$(ET)$_2$X (X=Cu[N(CN)$_2$]Br, Cu[N(CN)$_2$]Cl and Cu(NCS)$_2$) \cite{RevLang,Stalcup,RevMuller,Su,Yon_cool_04,Yon_cool_05}. 
Different studies of the hydrogenated $\kappa$(ET)$_2$Cu[N(CN)$_2$]Br and the deuterated analog $\kappa$(d$_8$-ET)$_2$Cu[N(CN)$_2$]Br have given evidence of inhomogeneous superconducting state whose volume fraction is found to be reduced by a quenching the sample \cite{Yon04,Sasaki05,Su,Stalcup,Kawamoto,Ito,830muller,831muller}.
Such inhomogeneity may explain the anomalous behavior of the inplane conductivity in these materials \cite{Singleton03}.
It has been also found that the inhomogeneous superconductivity in $\kappa$(ET)$_2$X compounds can also be generated by chemical \cite{Sasaki05,Kawamoto,Yon07,Tokumoto,Ito91,Miyagawa,Yon04} or hydrostatic pressure \cite{Sushko,Lefebvre,11muller,Kagawa,Muller} and irradiation \cite{Analytis,Sasaki10,James2}.\\

The outcome of the experimental studies on $\kappa$(ET)$_2$X and (TMTSF)$_2$X materials is that the proximity of the superconducting state to a magnetic insulating phase is a key ingredient for the formation of an inhomogeneous phase.
In the following, we present a theoretical approach to account for the dependence of the superconducting transition temperature on the disorder amount regardless of its origin. We assume that the disorder d\oe s not introduce local magnetic impurities in the ordered domains which is consistent with experimental findings.

\section{The model}

In this section, we first give a brief review of the theoretical approaches proposed in the literature to explain the behavior of T$_c$ as a function of disorder in organic conductors. We then present the outlines of the model we derived in Ref.\cite{EPL} to qualitatively interpret the dependence of T$_c$ on the cooling rate in (TMTSF)$_2$ClO$_4$. The key point lacking in Ref.\cite{EPL} is the correspondence between the theoretical parameters and the experimental disorder texture necessary to provide a quantitative comparison with the experimental results. In the remainder of the section, we propose a method the extract the values of these parameters from the experimental data.

\subsection{Theoretical approaches: brief review}

Powell and McKenzie \cite{Powell} have analyzed, in the framework of the AG law, the experimental results dealing with the effect of different types of disorder on the superconducting transition temperature T$_c$ of quasi-2D organic conductors.
It is worthwhile reminding that the AG law describes the suppression of T$_c$ by magnetic impurities for s-wave pairing. For non-s-wave superconductors, T$_c$ is suppressed by non-magnetic impurities according also to the AG formula given by \cite{AG}:
\begin{eqnarray}
\ln \frac{T_c}{T_{c0}}=\psi\left( \frac 12\right)
 -\psi\left( \frac 12+\frac{\hbar}{4\pi k_bT_c\tau}\right)
\label{AG}
\end{eqnarray}
where $T_{c0}$ is the superconducting critical temperature in the pure limit and $\psi(x)$ is the digamma function. $\tau$ is the quasi-particle lifetime due to scattering with magnetic impurities ($\tau=\tau_M$) or non-magnetic impurities ($\tau=\tau_N$).
Powell and McKenzie \cite{Powell} compared the experimental data obtained in $\beta$(ET)$_2$X and $\kappa$(ET)$_2$X with various types of disorder to the expectations of the AG formula. They found an excellent agreement using two free parameters to fit the theory. However, their study was not conclusive concerning the conventional nature of the superconducting order parameter, since the agreement with AG law holds for both magnetic and non-magnetic impurities. 
We emphasize that the data analyzed by Powell and McKenzie are in the clean limit corresponding to small amounts of disorder which acts as scattering points.
It is worth noting that Joo {\it et al.} \cite{Joo05} have also reported that the suppression of T$_c$ with increasing impurity concentration $x$ in the solid solution 
(TMTSF)$_2$(ClO$_4$)$_{1-x}$(ReO$_4$)$_{x}$ is consistent with AG formula. Moreover, based on EPR measurements, the authors \cite{Joo05} ruled out the magnetic nature of the scattering centers. The sensitivity of T$_c$ to the impurity concentration was, therefore, taken as a signature of the unconventional character of the superconducting gap in (TMTSF)$_2$ClO$_4$.
Despite the good agreement between the AG law and the experimental results in the clean limit of organic conductors, a clear discrepancy has been reported in the dirty limit of quasi-2D salts \cite{Analytis,Yon_cool_04,Yon04}.
To account for this discrepancy, Analytis {\it et al.}\cite{Analytis} suggested to use a generalized AG formula involving two order parameters with both s-wave and unconventional components. The presence of two gradients in the impurity induced suppression of T$_c$ was ascribed to this mixed superconducting ordering. The rapid decrease of T$_c$ for small disorder amount is attributed to the non-s-wave part whereas the slowing down of the T$_c$ suppression is assigned to the s-wave component. However, to the best of our knowledge, there is no experimental evidence of such mixed superconducting gap in organic conductors \cite{Powell}.\\

\subsection{Time Dependent Ginzburg-Landau approach}

Recently, we have proposed a model to describe the suppression of the superconducting transition temperature with increasing cooling rate in (TMTSF)$_2$ClO$_4$ \cite{EPL}. Basically, the model concerns a layered superconductor, where each layer, corresponding to the most conducting plane, has a phase segregation structure with superconducting domains embedded in a non-superconducting host. For simplicity, we assume that the superconducting clusters are identical and form a square lattice. These superconducting regions are interacting via Josephson couplings parameterized by $J_1$ and $J_2$ along the inplane directions $a$ and $b$ respectively. These domains are also coupled along the transverse direction $c$ by the interplane Josephson coupling $J_0$. This structure is consistent with the one proposed by M\"{u}ller {\it et al.} \cite{Muller} from fluctuation spectroscopy measurements in $\kappa$(ET)$_2$Cu[N(CN)$_2$]Cl. The authors have also argued that the mixed state of superconducting and non-superconducting regions can be regarded as a network of Josephson coupled junctions connecting the superconducting clusters.\

However, one should emphasize that in real samples the superconducting domains have not necessary planar structure with identical shape and size. For example, X ray measurements in (TMTSF)$_2$ClO$_4$ \cite{Pouget} suggested ellipsoidal structure. One should also consider randomly distributed superconducting clusters instead of the square lattice assumed in our model.
Actually, taking into account the thickness of the superconducting domain along the least conducting axis will not change the overall behavior of our results since superconductivity in the considered materials has essentially a two-dimension (2D) character.
Moreover, the shape of the superconducting domains is not a key factor in our model as far as the corresponding size is larger than the coherence length to avoid the phase fluctuations complexity. The basic parameter is the size of the non-superconducting junction on which depend the Josephson couplings.
It should be interesting to study the effect of a random distribution of the superconducting clusters which may be introduced by assuming a spatial weight dependence of the inplane Josephson coupling parameters. However, such correction is not expected to affect the outcomes of the present model. Indeed, the results depend on parameters extracted from experimental data which can be regarded as mean values associated to the randomly distributed superconducting islands in the real material.
It turns out, that considering different sized superconducting clusters with a random distribution may improve the quantitative agreement between our results and experiments but does not affect the general behavior.

In (TMTSF)$_2$ClO$_4$ and $\kappa$(ET)$_2$X salts, the Josephson junction is basically antiferromagnetic. Josephson effects through magnetic junctions have been discussed by Andersen {\it et al.} \cite{Andersen}.\

The anisotropic structure of the organic superconductors is at the origin of strong superconducting fluctuations which are considerably enhanced as the dimensionality of the system is reduced \cite{Varlamov}. Superconducting fluctuations, which can be brought out in many physical properties, express the presence of superconducting precursor effects in the normal state far from the superconducting transition temperature. These fluctuations smear out the sharp superconducting phase transition anomaly observed in 3D isotropic superconductors. In anisotropic systems, this transition is broadened and spread out over a transient regime where superconducting fluctuations take place \cite{Varlamov}.
Several experimental studies have given evidence for the presence of superconducting fluctuations in organic conductors \cite{RevMuller,Kwok,Nam}.

The critical superconducting fluctuations occur in the vicinity of $T_c$ over a thermal interval $\delta T$ characterized by large fluctuations of the order parameter.
The strength of the critical fluctuations is given by the so-called Ginzburg number \cite{Varlamov}:
\begin{eqnarray}
G_i=\frac{\delta T} {T_c}=\frac 1 2 \left[ \frac
{8\pi^2 k_B T_c \lambda(0)^2 \gamma}{ {\phi_0}^2 \xi_{\parallel} }
\right]^2
\end{eqnarray}
where the anisotropic parameter $\gamma=\frac{\xi_{\parallel}}{\xi_{\perp}}$ is the ratio of the inplane coherence length over the out-of-plane one, $\lambda(0)$ is the zero temperature inplane penetration depth and $\phi_0$  is the quantum flux.
For conventional superconductors $G_i$ is about $10^{-8}$ \cite{RevLang} whereas for hight-$T_c$ cuprates, $G_i$ amounts to $10^{-1}$ \cite{RevLang}. For (ET)$_2$X and (TMTSF)$_2$X salts, $G_i$ is of the order of $10^{-2}$ \cite{RevMuller,RevLang}. 
Due to the superconducting fluctuations, T$_c$ is reduced compared to the mean-field value $T_0$ obtained in a bulky superconductor in the absence of fluctuations \cite{Varlamov}.
$T_0$ can be estimated as $T_0=T^{exp}_c+G_i \,T^{exp}_c$ where $T^{exp}_c$ is the experimental critical temperature obtained in the pure limit.

The critical dynamics of the superconducting order parameter is governed by the Time Dependent Ginzburg-Landau (TDGL) equation \cite{Puica}
\begin{eqnarray}
\Gamma^{-1}_0\frac{\partial \psi_{nij}}{\partial t}=-
\frac{\partial F}{\partial\psi^{\ast}_{nij}}+\zeta_{nij}(\vec{r},t),
\label{partial}
\end{eqnarray}
where $\psi_{nij}$ is the superconducting order parameter in the ($i,j$) domain of the $n^{th}$ plane. Here $i$ ($j$) labels the superconducting island along the $a$ ($b$) direction. $\Gamma^{-1}_0=\pi{\hbar}^3/16 m \xi^2_0 k_BT$ is the relaxation rate of the order parameter, $\xi_0$ is the inplane coherence length and $m$ is the effective pair mass in the ($ab$) plane. The Langevin forces $\zeta_{nij}$ in Eq.\ref{partial} govern the thermodynamic fluctuations of the superconducting order parameter. $F$ is the free energy of the superconducting phase ($F_s$) compared to the normal state ($F_{norm}$) which is given by \cite{EPL}:
\begin{widetext}
\begin{eqnarray}
F=F_s-F_{norm}
&=&\sum_{i,j,n}\int_0^{L_1}dX\int_0^{L_2}dY \left[a|\psi_{nij}|^2+\frac{{\hbar}^2}{2m}|\vec{\nabla}\psi_{nij}|^2
\right.
+J_1|\psi_{nij}-\psi_{n\,i+1\,j}|^2\nonumber\\
&+&J_2|\psi_{nij}-\psi_{n\,i\,j+1}|^2
+\left.J_0|\psi_{nij}-\psi_{n+1\,i\,j}|^2
+\frac b 2 |\psi_{nij}|^4\right],
\label{free}
\end{eqnarray}
\end{widetext} 
The coefficients $a$ and $b$ are given by $a=a_0\epsilon$ and $b=\mu_0\kappa^2e_0^2{\hbar}^2/2m^2$, where $a_0={\hbar}^2/2m\xi^2_0$ and $\epsilon$ is the reduced temperature $\epsilon=\ln(T/T_0)$, $\kappa=\frac{\lambda_{\parallel}}{\xi_{\parallel}}$ is the GL parameter, $e_0=2e$ is the pair electric charge.
$L_1$ ($L_2$) is the size of the superconducting domain along the a (b) axis.
The Josephson couplings are
\begin{equation}
J_1=\frac{{\hbar}^2}{2m^{\ast}d_1^2},\quad \quad
J_2=\frac{{\hbar}^2}{2m^{\ast}d_2^2},\quad \quad \mbox{and}
\quad J_0=\frac{{\hbar}^2}{2m_c s^2}.
\label{J}
\end{equation}
Here $d_1$ ($d_2$) is the distance between two next neighboring superconducting domains along the $a$ ($b$) axis, while $s$ is the interplane distance(Fig.1).
$m^{\ast}$ and $m_c$ are the effective pair masses in, respectively, the superconducting domain and along the transverse $c$ axis. Hereafter, we consider $m^{\ast}=m$ since they correspond to the inplane effective Cooper pair masses whereas $m_c$ will be expressed as a function of the anisotropy parameter $\gamma$ as: $\gamma=\xi_0/\xi_c=\sqrt{m_c/m}$.
\begin{figure}[hpbt] 
\begin{center}
\includegraphics[width=1.\columnwidth]{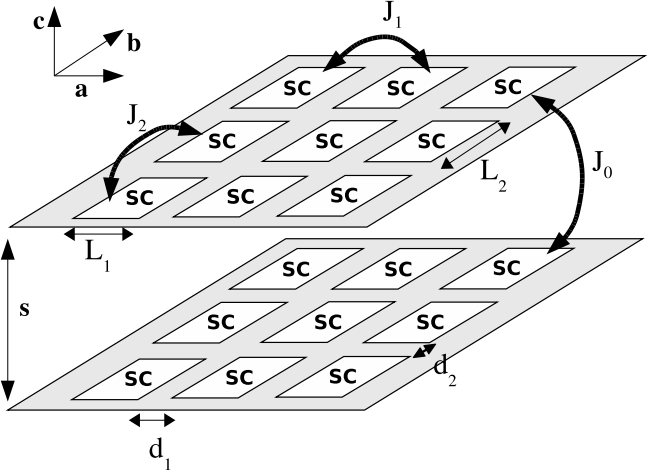}
\end{center}
\caption{Geometry of the inhomogeneous superconducting structure adopted in the model. The superconducting islands (denoted SC) are embedded in a non-superconducting matrix. We assume that the superconducting regions form an array for simplicity. The size of the superconducting domains is $L_1$ ($L_2$) in the $a$ ($b$) direction. The thickness of the non-superconducting domain is denoted by $d_1$ ($d_2$) along the $a$ ($b$) axis. The interplane distance $s$ has been exaggerated for clarity. The intraplane ($J_1$ and $J_2$) and the interplane ($J_0$) Josephson couplings are also represented.} 
\label{fig1}
\end{figure}
 
The Langevin forces $\zeta_{nij}$ are correlated through the Gaussian white-noise law \cite{Puica}:
\begin{eqnarray}
\langle \zeta_{nij}(\vec{r},t) \zeta^{\ast}_{n^{\prime}i^{\prime}j^{\prime}}(\vec{r}\;^{\prime},t^{\prime})
\rangle=2
\Gamma^{-1}_0k_BT\delta(\vec{r}-\vec{r}\;^{\prime})\delta(t-t^{\prime})\frac{\delta_{nn^{\prime}}}s
\end{eqnarray}
with $\vec{r}=(X+i(L_1+d_1),Y+j(L_2+d_2),ns)$
and $\vec{r}\;^{\prime}=X+i^{\prime}(L_1+d_1),Y+j^{\prime}(L_2+d_2),n^{\prime}s)$.

Assuming a Hartree approximation for the quartic term in Eq.\ref{free} leads to a linear problem where the $b|\psi_{nij}|^2\psi_{nij}$  term in the derivative of the free energy is replaced by $b\langle|\psi_{nij}|^2\rangle \psi_{nij}$ \cite{Puica}.
Given the Eqs. \ref{partial} and \ref{free}, the TDGL equation can be written as:
\begin{eqnarray} 
&&\zeta_{nij}(\vec{r},t)=\Gamma^{-1}_0\frac{\partial \psi_{nij}}{\partial t}-\frac{{\hbar}^2}{2m}\Delta\psi_{nij}+a \psi_{nij}\nonumber\\
&+&b\langle|\psi_{nij}|^2\rangle \psi_{nij}+J_1\left(2\psi_{nij}-\psi_{n\,i+1\,j}-\psi_{n\,i-1\,j}\right)\nonumber\\
&+&J_2\left(2\psi_{nij}-\psi_{n\,i\,j+1}-\psi_{n\,i\,j-1}\right)\nonumber\\
&+&J_0\left(2\psi_{nij}-\psi_{n+1\,i\,j}-\psi_{n-1\,i\,j}\right)
\label{zeta}
\end{eqnarray}

It is worth stressing that the TDGL theory we used was first proposed by Puica and Lang \cite{Puica} to study the effect of the superconducting fluctuations on the inplane conductivity in a self consistent Hartree approximation and by Mishonov {\it et al.}\cite{Mishonov} within a Gaussian approximation. Moreover, the non-Ohmic effect of the high electric field on the transverse magnetoconductivity has been studied in the frame of TDGL by Puica and Lang \cite{Puica06}. The authors have also interpreted, within the same approach, the excess Hall conductivity in the layered superconductors for an arbitrarily inplane electric field \cite{Puica09}. The TDGL has proven to be a powerful tool to investigate the role of superconducting fluctuations in the critical region to go beyond the usual Aslamazov-Larkin and Lawrence-Doniach approaches.\\

The form of the free energy given by Eq.\ref{free} is based on the model proposed by Puica and Lang \cite{Puica} to study the role of critical fluctuations on the conductivity of high-$T_c$ superconductors. The authors considered a layered structure with only interplane Josephson coupling $J_0$ connecting homogeneous superconducting layers.
Following the approach of Puica and Lang \cite{Puica}, the critical temperature T$_c$ of the superconducting state whose free energy is given by Eq.\ref{free} can be derived by solving the equation 
\begin{eqnarray} 
\tilde{\epsilon}=\epsilon+\frac b a \langle|\psi_{nij}|^2\rangle.
\label{self}
\end{eqnarray}
To solve this equation, we use the Green function method \cite{Puica} and define the Fourier transforms of $\psi_{nij}$ and $\zeta_{nij}(\vec{r},t)$ by:
\begin{widetext}
\begin{eqnarray*}
\psi_{nij}(x,y,t)=\int\frac{d^2\vec{k}}{(2\pi)^2}\int_{-\frac \pi s}^{\frac \pi s} \frac{dq}{2\pi}\psi_q(k_x,k_y,t){\rm e}^{-ik_x x}\;{\rm e}^{-ik_y y}\;{\rm e}^{-iqns}
\end{eqnarray*}

\begin{eqnarray}
\zeta_q(k_x,k_y,t)=\left[\Gamma^{-1}_0\frac{\partial }{\partial t}
+\frac{{\hbar}^2k^2_x}{2m}+\frac{{\hbar}^2k^2_y}{2m}+\tilde{a}
+2J_1\left(1-\cos(k_x(L_1+d_1)\right)
+2J_2\left(1-\cos(k_y(L_2+d_2)\right)
+2J_0\left(1-\cos(qs)\right)\right]\psi_q(\vec{k},t)
\label{Fzeta}
\end{eqnarray}
\end{widetext}
with
\begin{widetext}
\begin{eqnarray*}
\langle\zeta_q(\vec{k},t)\zeta^{\ast}_{q^{\prime}}(\vec{k}^{\prime},t^{\prime})\rangle=2\Gamma^{-1}_0 k_bT(2\pi)^3\delta(\vec{k}-\vec{k}^{\prime})
\delta(q-q^{\prime})\delta(t-t^{\prime})
\end{eqnarray*}
\end{widetext}

where $\vec{k}=(k_x,k_y)$ and $\tilde{a}=a+b\langle|\psi_{nij}|^2\rangle$.\

Equation \ref{Fzeta} is obtained by taking the Fourier transform of Eq.\ref{zeta}.\

We define the Green function $R_q(\vec{k},t,k^{\prime}_x,t^{\prime})$ of Eq.\ref{Fzeta} by:

\begin{widetext}
\begin{eqnarray}
\left[\Gamma^{-1}_0\frac{\partial }{\partial t}
+\frac{{\hbar}^2k^2_x}{2m}+a_1+ 2J_1(1-\cos(k_x(L_1+d_1))\right]R_q(\vec{k},t,k^{\prime}_x,t^{\prime})=
\delta(k_x-k^{\prime}_x)\delta(t-t^{\prime})
\label{R}
\end{eqnarray}
\end{widetext}
where 
\[
a_1=\tilde{a}+\frac{{\hbar}^2k^2_y}{2m}+2J_2(1-\cos(k_y(L_2+d_2))
+2J_0(1-\cos(qs). 
\]
The solution of Eq.\ref{Fzeta} is given by:
\begin{eqnarray*}
\psi_q(\vec{k},t)=\int dt^{\prime}\int dk^{\prime} _x R_q(\vec{k},t,k^{\prime}_x,t^{\prime})\zeta_q(k^{\prime}_x,k_y,t^{\prime})
\end{eqnarray*}

$\psi_q(\vec{k},t)$ can be derived by taking the Fourier transform of the Green function with respect to time \cite{Puica}

\begin{eqnarray*}
R_q(\vec{k},\omega,k^{\prime}_x,t^{\prime})=\int dt  R_q(\vec{k},t,k^{\prime}_x,t^{\prime})\rm{e}^{i\omega (t-t^{\prime})}
\end{eqnarray*}
Eq.\ref{R} can be written, then, as:
\begin{widetext}
\begin{eqnarray}
\left[-i\omega \Gamma^{-1}_0
+\frac{{\hbar}^2k^2_x}{2m}+a_1+2J_1(1-\cos(k_y(L_1+d_1))\right]R_q(\vec{k},t,k^{\prime}_x,t^{\prime})=
\delta(k_x-k^{\prime}_x)
\label{FR}
\end{eqnarray}
\end{widetext}
One can then obtain $\langle|\psi_{nij}|^2\rangle$ and solve Eq.\ref{self} for $\tilde{\epsilon}=0$ to derive the transition temperature T$_c$.
Straightforward calculations give rise to the following equation to which T$_c$ obeys:
\begin{widetext}
\begin{eqnarray} 
\ln \frac{T_c}{T_0}&+&\frac{8k_BT_cg}{\pi s}\int^{2\pi}_0d\theta\int^{E_c}_0 dE\int^{\pi}_{-\pi}d\varphi
\int^{w_c}_0dw
\left\lbrace w^2+\left[ E+2J^{\prime}_0\left( 1-\cos \varphi\right)\right.\right.\nonumber\\
&+&2J^{\prime}_1\left( 1-\cos \left( \frac{L_1+d_1}{\xi_{\parallel}}\sqrt{E}\cos \theta\right)\right.
+\left.\left. 2J^{\prime}_2\left( 1-\cos \left( \frac{L_2+d_2}{\xi_{\parallel}}\sqrt{E}\sin \theta\right)\right.\right]^2\right\rbrace =0
\label{Tc}
\end{eqnarray}
\end{widetext}
where $E_c$ and $w_c$ are cutoff parameters which depend on the energy scales of the system while $g$ and the reduced Josephson couplings $J^{\prime}_i$ ($i=0,1,2$) are given by:
$$
g=\frac{\mu_0\kappa^2e^2_0\pi\xi^2_{\parallel}}{8\hbar ^2} \quad \mathrm{and}
\quad J^{\prime}_i=\frac{J_i}{a_0}.
$$

From Eq.\ref{Tc}, one can derive the dependence of T$_c$ as a function of the $J^{\prime}_i$ couplings (Fig.1 of Ref.\cite{EPL}). However, a quantitative comparison with the experimental results requires the determination of the relationship between the Josephson couplings and the segregated structure induced by disorder. This point is discussed in the next section.\\

Let us now turn to the effect of the phase segregation on the upper critical field of Bechgaard salts. We focus, for simplicity, on the behavior of the upper critical field along the most conducting axis ($H_{c2}\parallel a$) to avoid complexity related to the field induced confinement occurring for $H\parallel b^{\prime}$. Moreover, we will tackle the case of (TMTFS)$_2$PF$_6$ in the vicinity of the critical pressure $P_c$ at which the SDW phase vanishes\cite{RevLebed}. In this region, the inhomogeneous structure is well established and the geometry of the domains has been determined experimentally \cite{PasquierPF6}.
We give in the following, the outlines of the model. Details will be published elswhere.\\

In the slab structure, reported in Refs. \cite{LeePF6,PasquierPF6}, superconducting domains of thickness $L$ are sandwiched by SDW insulating regions along the most conducting axis $a$.
We denote by $d$ the size of these SDW domains.
The free energy of the system in the presence of a magnetic field along the $a$ direction $\vec{H}=(H,0,0)$ is given by:
\begin{widetext}
\begin{eqnarray}
F=F_s-F_{norm}
=\sum_{n}\int dX\int dy\int dz \left[ a|\psi_{n}|^2+\frac{{\hbar}^2}{2m}|\left(\vec{\nabla}-i\frac{e_0}{\hbar} H y\right)
\psi_{n}|^2+J|\psi_{n}-\psi_{n+1}|^2+\frac b 2 |\psi_{n}|^4\right],
\label{free2}
\end{eqnarray}
\end{widetext}
where $\psi_{n}$ is the superconducting order parameter in the $n^{th}$ slab and $J$ is the Josephson coupling constant between two neighboring slabs: $J=\frac{{\hbar}^2}{2m^{\ast}d^2}$.\

Considering the gauge $\vec{A}=(0,0,H y)$, the TDGL equation becomes:

\begin{widetext}
\begin{eqnarray}
\Gamma_0^{-1}\frac{\partial \psi_n}{\partial t}+ a\psi_n+b|\psi_n|^2\psi_n+J\left(2\psi_n-\psi_{n-1}-\psi_{n+1}\right)
-\frac{\hbar^2}{2m}\left[\partial^2_x+\partial^2_y+\left(\partial_z-i\frac{e_0}{\hbar}Hy\right)^2\right]\psi_n=\zeta_n(\vec{r},t)
\end{eqnarray}
\end{widetext}
Using the method introduced by Ullah and Dorsey \cite{Ullah} and by Puica and Lang\cite{PuicaH} in the case of layered superconductors where superconducting planes are coupled via Josephson parameters, one can deduce the upper critical field $H_{c2}^a$ for a superconductor with a slab structure. Details of calculations will be given in a forthcoming paper. The results is discussed in the next.

\subsection{Lessons from experiments}
Disorder can be generated, as we have seen, by different sources: cooling, anion or donor substitution and irradiation. Regardless of the origin of the introduced disorder, we have to establish the connection between the experimental parameters measuring its amount and the size of the non-superconducting domain on which depend the Josephson couplings $J_1$ and $J_2$ given by Eq.\ref{J}.
For simplicity, we assume hereafter that the superconducting clusters are isotropic with a length $L$. This assumption d\oe s not affect the outcomes of the study as we shall show in the following. The size of the non-superconducting domain is denoted, henceforth, $d$.

\subsubsection{The case of (TMTSF)$_2$ClO$_4$}

Pouget {\it et al.} \cite{Pouget} have carried out a high resolution structural study on (TMTSF)$_2$ClO$_4$ based on X-ray synchrotron radiation diffraction. The authors determined, for different cooling rates, the average size $L_i$ (i= $a$, $b$, $c$) along the crystallographic directions and the fraction $f$ of the ordered ClO$_4$ domains where superconductivity develops at low temperature.
From these data, we have derived the typical size of the non-ordered domains assuming that the ordered domains have a cubic shape with a length $L$ separated by non-superconducting slabs of thickness $d$ along the three directions.
This 3D structure seems to be in contradiction with the assumption of a layered superconductor considered in our model.
Actually, the superconducting phase in organic conductors has a 3D character but regarding the relatively small coherence length $\xi_c$ along the least conducting axis $c$, this phase is substantially dominated by the inplane parameters. The coherence lengths in (TMTSF)$_2$ClO$_4$ along the crystallographic axes are typically of $\xi_a\sim 800 \AA$, $\xi_b\sim 300 \AA$ and $\xi_c\sim 20 \AA$ \cite{Revue,Pouget}.\newline
$\xi_c$ amounts to about the interplane distance $c=13 \AA$ (denoted $s$ in Eq.\ref{J}) which supports the assumption of an interlayer Josephson coupling whatever the cooling rate.\

Considering the isotropic structure, the fraction of the ClO$_4$ ordered domains can be written as:
\begin{equation}
 f=\left (\frac L{L+d}\right)^3
\label{f}
\end{equation}
Given the experimental data for $f$ and $L$, one can deduce the values of $d$ in order to have a quantitative comparison between our theoretical results and experiments. In Table 1 we have listed, for different cooling rates, the experimental values of the superconducting volume fraction $f$, the lengths of the ordered domains along the crystallographic axes denoted $L_a$, $L_b$ and $L_c$ and the average length $L=(L_aL_bL_c)^{\frac 13}$ of a superconducting domain within the isotropic assumption. We also give the typical size $d$ of the non-superconducting domain and the renormalized Josephson coupling  , where $J^{\prime}=\frac J{a_0}=\left(\frac {\xi_0}d\right)^2$ (Eqs.\ref{free},\ref{J}). These values are calculated for an average inplane coherence length $\xi_0=\sqrt{\xi_a\xi_b}=500 \AA$.
\begin{widetext}
\begin{center}
\begin{table}[htbp]
\begin{center}
\begin{tabular}{|c|c|c|c|c|c|c|c|c|}
\hline
Cooling rate (K/min)& f & $L_a$ ($\AA$)& $L_b$($\AA$)& $L_c$($\AA$)& $L=(L_aL_bL_c)^{\frac 13}$($\AA$)& $d$($\AA$)& $\frac d {\xi_0}$ x$10^2$&$J^{\prime}$\\
\hline
0.5 & 0.95& 700 &1050&1150& 950&16&3.2&976.5\\
\hline
1& 0.9 & 550& 900& 900& 750& 27&5.4&342.9\\
\hline
3 & 0.7& 400& 650& 550& 500 & 63&12.6&62.9\\
\hline
5 & 0.48& 300& 550& 450 & 420& 116&23.2&18.6\\
\hline
\end{tabular}
\end{center}
\vspace{0.5pt}

\label{Table1}
\tablename {  1:
Superconducting fraction $f$, typical sizes $L_i$ and the corresponding average $L$ value of a superconducting domain in (TMTSF)$_2$ClO$_4$ at different cooling rates. The interdomain distance $d$ and the renormalized Josephson coupling $J^{\prime}$ are also indicated (data of Ref.\cite{Pouget}).}

\end{table}
\end{center}
\end{widetext}
\vspace{0.3cm}

In our model, the average size $L$ of the superconducting domain is assumed to be larger than the coherence length $\xi_0$ to avoid any complexity related to phase fluctuations of the superconducting order parameter. According to the data of table 1, this assumption breaks down for a cooling rate greater than 3 K/min for the average inplane coherence length $\xi_0\sim 500 \AA$.
Actually, for such rates, the Josephson coupling is no more efficient regarding the relatively large size of the non-superconducting junction. It then reduces to a tunneling effect. At constant Josephson coupling, the superconducting transition temperature tends to saturation as far as the size $L$ of the superconducting domain is larger than the inplane coherence length $\xi_0$. Indeed, numerical results \cite{Mannai_unpb} have shown that, in this case, the superconducting transition temperature is practically independent of $L$. However, for rapid cooling rate, experimental data \cite{Pouget} show that the size $d$ of the Josephson junction is only slightly changed whereas the superconducting average domain size $L$ is strongly reduced giving rise to the collapse of the superconducting transition temperature.
In the present work, we will focus on the slow and intermediate cooling rates for which $T_c$ exhibit a clear departure from the AG law.
We will then assume that $L$ is roughly unchanged by cooling and remains larger than $\xi_0$.
Despite this simplifying assumption, we will show in the next section, that the obtained results are quantitatively in agreement with the experimental ones.

\subsubsection{The case of $\kappa$(ET)$_2$X salts}

In table 2, we list the different symbols used in the text.\\

\begin{widetext}
\begin{center}
\begin{table}[h]
\begin{center}
\begin{tabular}{|c|c|}
\hline
Symbol & Definition \\
\hline
$L$ & Size of the superconducting domain\\
\hline
$d$ & Size of the non-superconducting domain\\
\hline
$f$ & Superconducting volume fraction\\
&$f=\left(\frac{L}{L+d}\right)^3$\\
\hline
$l_{\parallel}$ & in-plane mean free path $^a$\\
\hline
$l_{\parallel}(0)$ & in-plane mean free path in the cleanest limit\\
& (smallest disorder rate)\\
\hline
$\rho_0^{\ast}$ & residual resistivity at which $T_c$ deviates from AG law\\
\hline
$l_{\parallel}^{\ast}$ & in-plane mean free path corresponding to $\rho_0^{\ast}$ \cite{Analytis}  \\
\hline
$d(0)$ & Size of the non-superconducting domain in the cleanest limit\\
\hline
$\delta$& Size increase of the non-superconducting domain by increasing disorder rate\\
& $d=d(0)+\delta$\\
\hline
\end{tabular}
\end{center}
\vspace{0.5cm}

\label{Table2}
\tablename { 2 :
Definition of the different symbols used in the text.\

$^a$ $l_{\parallel}$ is taken equal to the size $L$ of the superconducting domain assuming that the latter is free of impurities. This assumption is based on the idea that we only consider the effect of inhomogeneous superconductivity induced by disorder and disregard any possible local impurities inside the superconducting island.}

\end{table}
\end{center}
\end{widetext}
\vspace{0.4cm}

The effect of the disorder induced by the cooling rate on $\kappa$(ET)$_2$X (X=Cu(NCS)$_2$ and Cu[[N(CN)$_2$]Br) has been investigated by Yoneyama {\it et al.} \cite{Yon_cool_04} by measuring the inplane penetration depth. The authors concluded that the local clean approximation is in good agreement with experimental results, indicating that disorder induced by fast cooling leads to a short inplane mean free path $l_{\parallel}$. The authors have also obtained, from the data of Stalcup {\it et al.} \cite{Stalcup}, the behavior of the superconducting transition temperature T$_c$ with the Dingle temperature T$_D$ which is an indicator of the sample purity regarding its dependence on the mean free path $l_{\parallel}$: $T_D=\frac{v_F\hbar}{2\pi k_B l_{\parallel}}$, where $v_F$ is the Fermi velocity.
It is worth stressing that the inplane mean free path measured from the intralayer component of the conductivity may be different from those deduced from cyclotron resonance experiments or de Haas-van Alphen effect as discussed by Singleton in Ref.\cite{SingletonRev}. It has also been found that the measured Dingle temperature contains the effect of spatial inhomogeneities \cite{SingletonRev}. \

The domains, where the ethylene endgroups are disordered by quenching, act as scattering centers according to the results of Yoneyama {\it et al.} \cite{Yon_cool_04}. The faster the cooling, the shorter the mean free path.
By quenching, T$_c$ is reduced whereas the penetration depth increases, in particular for the Cu[[N(CN)$_2$]Br salt. One can then describe the system as a bulk superconductor where the number of defect centers increases with cooling rate. These defects are actually non-superconducting domains of a size $d$ where the ethylene groups are expected to be disordered.
Actually, the idea attributing the disorder in $\kappa$-(ET)$_2$ salts to the quenching ethylene endgroups disorder has been widely supported \cite{RevMuller}. This idea is not, however, corroborated by the structural investigation reported by Wolter {\it et al.} \cite{Wolter} who have argued that the configurationally disorder of ethylene endgroups is not the unique factor. The detailed structural study of Wolter{\it et al.} \cite{Wolter} is in agreement with the results of Strack {\it et al.} \cite{Strack} showing that the contribution of the frozen-in disorder of the ethylene group is not substantial.

It is worth noting that the defects induced by the configurational disorder of the ethylene endgroups cannot be considered as local impurities since the network of $H$-bondings of the anions with the disordered terminal ethylene groups induce a displacive disorder involving several ET molecules \cite{Pouget93}.
From this picture, the mean free path $l_{\parallel}$ can be taken as the size of the superconducting domain in our isotropic model.
Let us denote by $l_{\parallel}(0)$ the mean free path associated to the cleanest limit and $d(0)$ the corresponding distance between neighboring superconducting domains.
At a given cooling rate, $l_{\parallel}$ writes as $l_{\parallel}=l_{\parallel}(0)-\delta$ and as a consequence $d=d(0)+\delta$ due to the isotropic structure of the superconducting cluster we assumed in our model.
The experimental values of $l_{\parallel}$ and $l_{\parallel}(0)$ can be deduced from the Dingle temperature (Fig.8 of Ref.\cite{Yon_cool_04}). By adjusting $l_{\parallel}(0)$ to obtain numerically the critical temperature T$_c$ for the slowest cooling rate, one can simply extract the experimental values of $d$ for different cooling rates and derive the theoretical dependence of T$_c$ as a function of the cooling rate or the Dingle temperature.\\

In table 3 we give the main parameters inferred from the experimental data of Ref.\cite{Yon_cool_04}.

\begin{widetext}
\begin{center}
\begin{table}[h]
\begin{center}
\begin{tabular}{|c|c|c|}
\hline
Dingle Temperature $T_D$ (K)& Mean free path $l_{\parallel}$($\AA$)& $d$ ($\AA$)\\
\hline
 2.04& 372.5 & 7.5 \\
\hline
2.25& 338&  42\\
\hline
2.5 & 304 & 76\\

\hline
\end{tabular}
\end{center}
\vspace{0.5cm}

\label{Table3}
\tablename { 3 :
Some values of the inplane mean free path $l_{\parallel}$ and the size $d$ of the non-superconducting domains deduced from the experimental values of the Dingle temperature which have been obtained by Yoneyama {\it et al.} \cite{Yon_cool_04} from the data of Stalcup {\it et al.}\cite{Stalcup}.
We have assumed that the cleanest limit corresponds to the highest transition temperature which is ascribed, according to the data of Ref.\cite{Yon_cool_04}, to $T_D$= 2 K.
The latter gives $l_{\parallel}(0)\sim$ 380 $\AA$.}
\end{table}
\end{center}
\end{widetext}
\vspace{0.3cm}

Yoneyama {\it et al.} \cite{Yon04} have studied the effect of deuterated ET molecule on the superconducting state of $\kappa$-[(h$_8$-ET)$_{1-x}$(d$_8$-ET)$_x$]$_2$Cu[N(CN)$_2$]Br where h$_8$-ET and d$_8$-ET correspond, respectively, to the fully hydrogenated and fully deuterated molecules. Moreover, Sasaki {\it et al.} \cite{Sasaki05} have obtained, based on scanning micro region infrared spectroscopy, the superconducting volume fraction $f$ in this salt as a function of the substitution ratio $x$.
We can, then, deduce, for each concentration $x$, the size $d$ of the non-superconducting domain as (Eq.\ref{f}): $\frac d L=\left(\frac 1 f\right)^{\frac 13}-1$. 
According to our calculations \cite{EPL,Mannai_unpb}, the superconducting transition temperature T$_c$ is, practically, unchanged by varying the size $L$ of the superconducting cluster. The latter should be, however, greater than the inplane coherence length $\xi_0$ to avoid, as we have previously discussed, any problem related to the phase coherence of the order parameter as it is the case in granular superconductors \cite{Granular}.

We take $L\sim 100 \AA$ whatever the $x$ value since for the considered $\kappa$(ET)$_2$X salts $\xi_0\sim 30\AA-70\AA$.
Given the experimental values of $f$ for different concentration $x$, one can, then, deduce $d$ and establish the correspondence between the Josephson couplings of our model and the substitution ratio $x$ to derive the theoretical behavior of T$_c$ as a function of $x$.\
It is worth stressing that we will consider the region of inhomogeneous superconductivity corresponding to $x$ in the range $0.5-1$ in the phase diagram of Yoneyama {\it et al.} \cite{Yon04}. The cleanest limit, corresponding to a homogeneous superconducting state, is ascribed to $x=0.5$ (Fig.5 of Ref.\cite{Yon04}).\\

In table 4 we give the main parameters deduced from the experimental data of Sasaki {\it et al.} \cite{Sasaki05}.

\begin{widetext}
\begin{center}
\begin{table}[h]
\begin{center}
\begin{tabular}{|c|c|c|}
\hline
$x$& superconducting & $\frac d L\times 10^2$\\
d$_8$-ET concentration & volume fraction $f$&\\
\hline
 0.5 & 1 & 0 \\
\hline
0.6& 0.9&  3.6\\
\hline
0.7 & 0.82 & 6.8\\
\hline
0.8 & 0.68 & 13.6\\
\hline
0.9 & 0.54 & 23\\
\hline
1& 0.44& 31.5\\
\hline
\end{tabular}
\end{center}
\vspace{0.5cm}

\label{Table4}
\tablename { 4 :
Values of the superconducting volume fraction $f$ and the ratio $\frac d L$ extracted from the experimental values of Ref.\cite{Sasaki05}. Here $\frac d L=\left(\frac 1 f\right)^{\frac 13}-1$. We have taken, as in Ref.\cite{Yon04}, the cleanest limit at $x=0.5$.}
\end{table}
\end{center}
\end{widetext}
\vspace{0.3cm}

The T$_c$ dependence on the irradiation damage has been studied by Analytis {\it et al.}\cite{Analytis} in $\kappa$(ET)$_2$Cu(NCS)$_2$. The defect density was probed by the residual resistivity $\rho_0$ which is proportional to the scattering rate $\tau$. 
We denote by $ \rho^{\ast}_0$ and $l^{\ast}_{\parallel}$ the critical value of $\rho_0$ and the corresponding mean free path above which T$_c$ d\oe s not follow a linear behavior \cite{Analytis}.
Analytis {\it et al.}\cite{Analytis} have found that for $ \rho^{\ast}_0$, $\frac{\xi_0}{l^{\ast}_{\parallel}}\sim 0.2$ where $\xi_0\sim 70\AA$ is the inplane coherence length. This leads to the estimation $l^{\ast}_{\parallel}\sim 350\AA$.
For a given irradiation amount, we can deduce the mean free path as $l_{\parallel}=l^{\ast}_{\parallel}\frac{\rho^{\ast}_0}{\rho_0}$ since $\rho_0$ is proportional to the inverse of the inplane mean free path $l_{\parallel}$.
It is worth noting that the determination of the the inplane scattering rate or the inplane mean free path from the interplane residual resistivity has been used by Joo {\it et al.} \cite{Joo04,Joo05}  in the case of the quasi-one dimensional organic superconductor (TMTSF)$_2$ClO$_4$ to fit, using the Abrikosov-Gor’kov formula, their experimental data dealing with the dependence of the superconducting transition temperature with the interplane residual resistivity.
The result of Joo {\it et al.} was taken as an evidence of the unconventional character of the superconducting order parameter in (TMTSF)$_2$ClO$_4$.
Moreover, the extraction of the inplane scattering rate and the inplane mean free path from interlayer resistance to fit Abrikosov-Gor’kov law has been also used by B. Powell and R. Mackenzie \cite{Powell} in the case of several quasi-two dimensional organic superconductors.\\

Following the method we introduced above, $l_{\parallel}$ can be written as $l_{\parallel}=l_{\parallel}(0)-\delta$, where $l_{\parallel}(0)$ is the mean free path in the clean limit corresponding to the highest T$_c$ obtained for $\rho_0\sim 0.33\,\Omega$cm. This gives $l_{\parallel}(0)\sim 6 \,l^{\ast}_{\parallel}$.
The distance $d$ between neighboring superconducting regions reads, then: $d=d(0)+\delta$, where $\delta=l_{\parallel}(0)-l_{\parallel}\sim l^{\ast}_{\parallel} \left[6-\frac{\rho^{\ast}_0}{\rho_0}\right]$.
$d(0)$ is taken from the fit of the T$_c$ values in the cleanest limit.

Table 5 shows the main parameters inferred from the experimental data of Analytis {\it et al.}\cite{Analytis}.

\begin{widetext}
\begin{center}
\begin{table}[h]
\begin{center}
\begin{tabular}{|c|c|c|}
\hline
$\rho_0$ ($\Omega$cm)& Mean free path $l_{\parallel}=l^{\ast}_{\parallel} \frac{\rho^{\ast}_0}{\rho_0}$ (in the unit of $l^{\ast}_{\parallel}$)& $\delta=l_{\parallel }(0)-l_{\parallel}$ (in the unit of $l^{\ast}_{\parallel}$)\\
\hline
 0.33 & 6 & 0 \\
\hline
0.66& 3&  3\\
\hline
1.5 & 1.3 & 4.7\\
\hline
2 & 1. & 5.\\
\hline
2.6 & 0.7 & 5.3\\
\hline
6& 0.3& 5.7\\
\hline
\end{tabular}
\end{center}
\vspace{0.5cm}

\label{Table5}
\tablename { 5 :
Some values of the mean free path $l_{\parallel}$ and the non-superconducting size increase $\delta$ in units of $l^{\ast}_{\parallel}$ corresponding to the value $\rho_0^{\ast}$ of the residual resistivity at which the superconducting transition temperature $T_c$ deviates from the AG law. According to the data of Analytis {\it et al.}\cite{Analytis},$\rho_0^{\ast}\sim 2 \Omega$cm. Taking a coherence length of $\xi_0\sim 70 \AA$ gives $l^{\ast}_{\parallel}\sim 350 \AA$. At the cleanest limit, corresponding to the highest $T_c$, we obtain $l_{\parallel}(0)\sim 6 l^{\ast}_{\parallel}$. }
\end{table}
\end{center}
\end{widetext}
\vspace{0.3cm}

\section{Results and discussion}

In figure 2 we plot the superconducting transition temperature T$_c$ obtained from Eq.\ref{Tc} as a function of the cooling rate for (TMTSF)$_2$ClO$_4$ salt. The results are in good agreement with the experimental data of Matsunaga {\it et al.}\cite{Mat} and Joo {\it et al.}\cite{JooPhD}. As shown in figure 2, T$_c$ falls with increasing cooling rate but presents a significant slope change at a cooling rate of about 5 K/min above which the effect of the cooling rate on $T_c$ is reduced. For slow cooling (less than 5 K/min), the tiny non-superconducting domains act as local defects which are known, according to the AG law, to have drastic effect on superconductivity. Since EPR measurements \cite{Tomic82,JooPhD} revealed the  non-magnetic nature of the cooling induced disorder centers in (TMTSF)$_2$ClO$_4$, the sensitivity of T$_c$ to the cooling rate is a signature of the non-s-wave character of the superconducting order parameter.
By quenching, the non-superconducting regions get wider and can no more be considered as point defects. The decrease of T$_c$ is slowed down and d\oe s no more follow the AG formula.\\
\vspace{1cm}
\begin{figure}[hpbt] 
\begin{center}
\includegraphics[width=0.9\columnwidth]{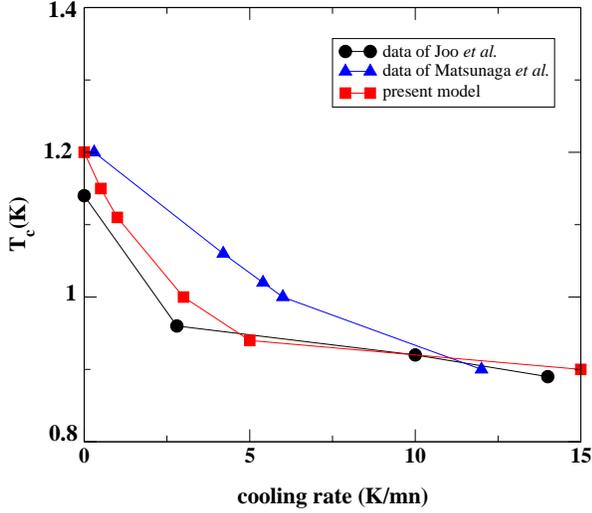}
\end{center}
\caption{Superconducting transition temperature as a function of the cooling rate in (TMTSF)$_2$ClO$_4$. Squares correspond to our results obtained from Eq.\ref{Tc} whereas the triangles and filled circles are, respectively, the data of Refs.\cite{Mat,JooPhD}. Lines are guide to eye.} 
\label{fig1}
\end{figure}

The tendency to saturation in this regime reflects the insensitivity of the Josephson coupling due to the increasing size of the Josephson junction between the superconducting domains. Actually, this saturating regime is expected to end as soon as the size of the superconducting domain becomes smaller than the inplane coherence length leading to the disappearance of the superconducting state. \
It is worth noting that, despite the simplifying assumption considered in our model, the obtained results show good agreement with experimental data. This indicates that actually, in the range of validity of the calculations, the cooling rate acts principally on the size of the non-superconducting domains. It turns out that the ratio $\frac d{\xi_0}$, on which depend the reduced Josephson coupling constants $J^{\prime}=\frac J {a_0}$ in Eq.\ref{Tc}, is the key parameter governing the dependence of the superconducting transition on the cooling rate.\\

In figure 3, we depicted the dependence of T$_c$ on the Dingle temperature corresponding to different cooling rates in $\kappa$-(ET)$_2$Cu[N(CN)$_2$]Br. The good agreement between our results and the experimental data \cite{Yon_cool_04} supports the method used to connect the microscopic parameters of our model with the experimental amount of disorder. It also reflects that the key ingredient is the size $d$ of the non-superconducting domains which governs the strength of the Josephson coupling.\\
\vspace{0.3cm}

\begin{figure}[htpb] 
\begin{center}
\includegraphics[width=0.9\columnwidth]{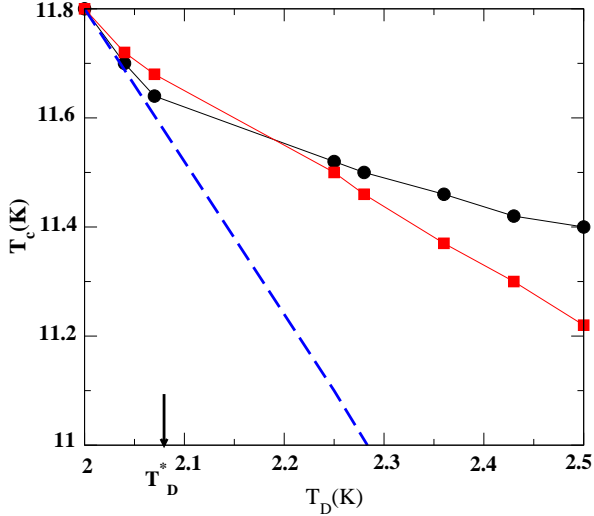}
\end{center}
\caption{Superconducting transition temperature as a function of Dingle temperature for $\kappa$-(ET)$_2$Cu[N(CN)$_2$]Br. Squares correspond to the results obtained from Eq.\ref{Tc} whereas the filled circles are the data of Ref.\cite{Yon_cool_04}. $T^{\ast}_D$ denotes the value of the Dingle temperature at which the experimental data deviate from the AG law.} 
\label{fig2}
\end{figure}

In figure 3 we also give the fit of the experimental data with AG formula. At low disorder amount (small $T_D$), the experimental results are well described by the AG law, indicating that the non-superconducting domains are tiny enough and can be considered as defect points. In this regime, our results are also in agreement with the AG formula which means that one recovers the AG law as a limiting case corresponding to small disorder amount.
However, for a large disorder amount, there is a clear discrepancy between the AG law and the experimental data since, in this case, the defect point picture d\oe s not hold anymore regarding the increasing size of the non-superconducting domains.
The critical Dingle temperature $T^{\ast}_D$, marked by an arrow in Fig.2, corresponds to a ratio $\frac {d^{\ast}}{\xi_0}\sim 0.2$. For $d>d^{\ast}$, the non-superconducting domains are large enough and the AG formula is no more justified.
From the value of $T^{\ast}_D$, one can also deduce the reduced critical value of the mean free path $l^{\ast}_{\parallel}$ : $\frac {\xi_0} {l^{\ast}_{\parallel}}\sim 0.17$.\\

Figure 4 (a) and (b) show the behavior of T$_c$ with the deuteration rate $x$ in $\kappa$-[(h$_8$-ET)$_{1-x}$(d$_8$-ET)$_x$]$_2$Cu[N(CN)$_2$]Br for the relaxed and quenched samples respectively. Our results are consistent with the data of Yoneyama {\it et al.} \cite{Yon04}. The calculated T$_c$ seems to have a decrease rate somewhat greater than the experimental values, in particular for the quenched sample. This may be due to the simplified assumption concerning the isotropic character of the superconducting domains which we used to derive the length of the non-superconducting regions from the experimental data.
With increasing $x$, the non-superconducting islands develop resulting in a reduced T$_C$. However, the effect of deuteration induced disorder appears to be rather negligible. The suppression of T$_c$ was ascribed by Yoneyama {\it et al.}\cite{Yon04} to the presence of magnetic impurities. The non-superconducting domains should be AF since the ground state of the d$_8$-ET salt ($x=1$) is AF with minor superconducting regions.

\begin{figure}[t] 
\begin{center}
\includegraphics[width=0.9\columnwidth]{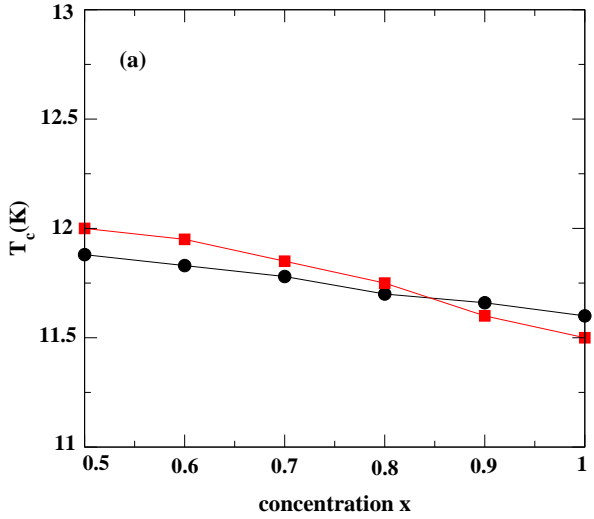}
\vspace{1.5cm}

\includegraphics[width=0.9\columnwidth]{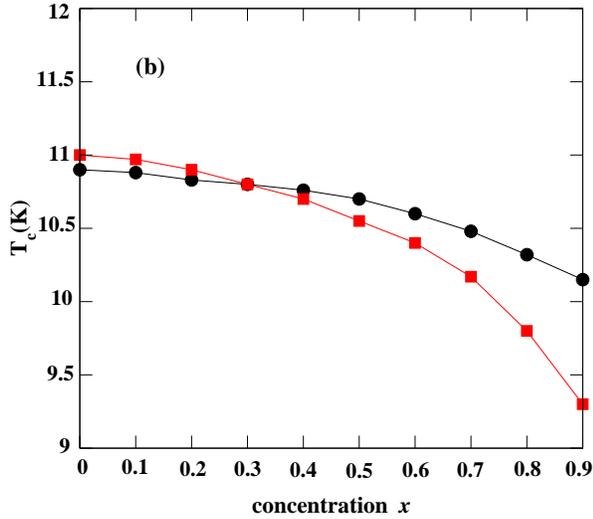}
\end{center}
\caption{Superconducting transition temperature as a function of the deuteration concentration $x$ for $\kappa$-[(h$_8$-ET)$_{1-x}$(d$_8$-ET)$_x$]$_2$Cu[N(CN)$_2$]Br in the relaxed (a) and quenched (b) samples. Squares correspond to the results obtained from Eq.\ref{Tc} and the filled circles are the data of Ref.\cite{Yon04}.} 
\label{fig3}
\end{figure}

The suppression of T$_c$ with irradiation in $\kappa$(ET)$_2$Cu(NCS)$_2$ is shown in Fig.5 where the residual resistivity $\rho_0$ measures the irradiation damage. A fit by the AG formula is indicated by the dashed line. Our results describe correctly the experimental data of Analytis {\it et al.} \cite{Analytis} and those of Sasaki {\it et al.} \cite{Sasaki10}, which corroborates the idea that irradiation generate non-superconducting grains in the bulk of the sample and not only at the surface.
Moreover, the departure from the AG law can be understood within the present model as a reduction of the Josephson tunneling due to increasing size of the non-superconducting regions. There is no need to assume a multicomponent superconducting order parameter as proposed by Analytis {\it et al.} \cite{Analytis}, especially since there is no experimental evidence for such scenario.
\begin{figure}[t] 
\begin{center}
\includegraphics[width=8cm,height=6cm]{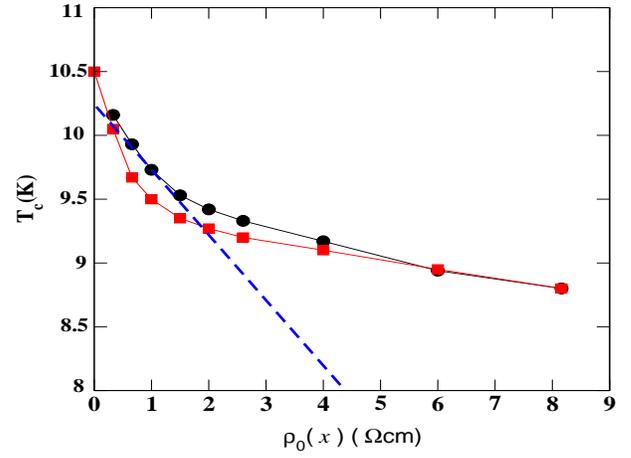}
\end{center}
\caption{Superconducting transition temperature as a function of the residual resistivity in $\kappa$(ET)$_2$Cu(NCS)$_2$. Squares correspond to the results obtained from Eq.\ref{Tc} and the filled circles are the data of Ref.\cite{Analytis}.} 
\label{fig4}
\end{figure}

At the critical value $\rho^{\ast}$ at which the data show a deviation from the AG law, Analytis {\it et al.} \cite{Analytis} have found that $\frac {\xi_0} {l^{\ast}_{\parallel}}\sim 0.2$. This is practically the same value we obtain in the case of $\kappa$-(ET)$_2$Cu[N(CN)$_2$]Br where $\frac {\xi_0} {l^{\ast}_{\parallel}}\sim 0.17$.
It seems that there is a critical ratio which limits the range of validity of the AG law in $\kappa$-(ET)$_2$X salts. More experimental data are needed to generalize this finding.\\

The outcome of these results is that our model, despite its simplicity, provides a coherent interpretation of the role of disorder in layered inhomogeneous superconductors. The technique used to extract the leading microscopic parameter of the model related to the non-superconducting domain length, has proved to be reasonable regarding the good agreement between the calculations and the experimental findings.\\

Figure 6 shows the behavior of the upper critical field along the $a$ axis for different rates of disorder assuming the slab structure found in the case of (TMTSF)$_2$PF$_6$. The parameter $d$ is the size of the non-superconducting slabs. In the case of large disorder amount, corresponding to a large $d$, $H_{c2}$ saturates at a value around the Pauli limit ($\sim 2.5$ T). However, by reducing the disorder concentration, which turns out to reduce $d$, $H_{c2}$ is greatly enhanced with a non saturating behavior as observed experimentally \cite{LeePF6}.
We ascribe this feature to the superconducting fluctuations which are enhanced by reducing $d$ giving rise to a robust superconducting state. 
Our results suggest that the superconducting fluctuations are responsible of the non-saturating behavior of $H_{c2}$ in (TMTSF)$2$PF$_6 $ salt. This finding may shed light on the origin of the discrepancy between the $H_{c2}$ values obtained by thermodynamic and transport measurements in (TMTSF)$2$ClO$_4$\cite{Shinagawa,Yonezawa_up,Yonezawa_P}.
The former reveal a saturating upper critical fields whereas the latter show a non saturating behavior. Transport measurements are sensitive to superconducting fluctuations. According to our results, the divergent character of the upper critical fields reported by transport measurements is a signature of superconducting fluctuations.\
Regarding the non saturating behavior of the upper critical field, the high field phase has been ascribed to a triplet state whereas the low field phase is assumed to be a singlet phase\cite{DeMelo}.\

It is worth to note, that, as done by Ullah and Dorsey \cite{Ullah} and Puica and Lang\cite{PuicaH}, we do not consider in our calculations a particular symmetry of the superconducting state. The model deals with unconventional superconductors regarding the dependence of the critical temperature on disorder.
However, according to our result, the high field phase could be a singlet state and the divergent character is nothing but the signature of the superconducting fluctuations.
\vspace{1cm}
\begin{figure}[hpbt] 
\begin{center}
\includegraphics[width=0.9\columnwidth]{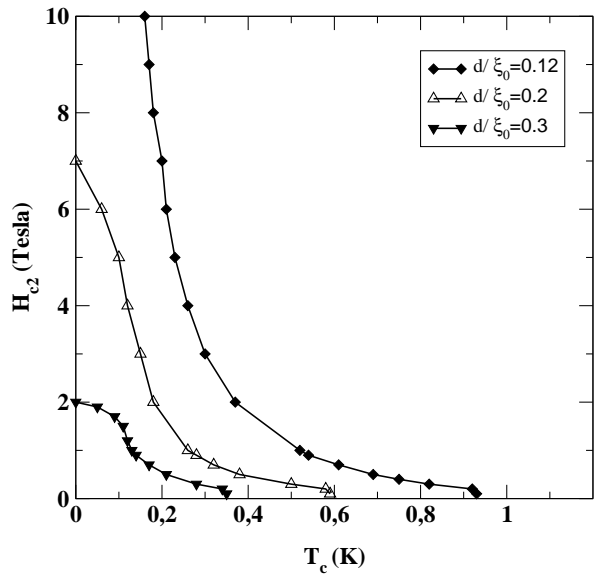}
\end{center}
\caption{Upper critical fields in layered superconductor with a slab structure. Calculations are done for (TMTSF)$_2$PF6 data. $d$ is the thickness of the non superconducting domains and $\xi_0$ is the coherence length ($\xi_0\sim 500 \AA$)}. 
\label{fig5}
\end{figure}
 
\section{Concluding remarks}

We have studied the effect of disorder, regardless of its origin, on the behavior of the superconducting critical temperature $T_c$ in the Bechgaard and  $\kappa$-(ET)$_2$X salts. We have considered an inhomogeneous superconducting phase where superconducting domains are embedded into a non-superconducting matrix.
For small disorder amount the latter can be considered as defect point and the suppression of $T_c$ is found to be well described by the AG law. However, in the regime of large disorder rate, the non-superconducting regions get wider and the results expected from AG formula are in clear discrepancy with experimental data.
We have then proposed a model to describe the suppression of $T_c$ taking into account the inhomogeneous structure of the disordered material. We have shown that this texture gives rise to Josephson tunneling which depends on the key parameter $\frac d{\xi_0}$ where $d$ is the size of the non-superconducting junction and $\xi_0$ is the inplane coherence length.
We have also found that, below a critical value $\eta^{\ast}$ of the ratio $\eta= \frac{\xi_0}{l_{\parallel}}$, where $l_{\parallel}$ is the inplane mean free path, the AG law holds. The non-superconducting regions act as impurities in this case.
$\eta^{\ast}$ is found to be of the order of 0.2 in $\kappa$ (ET)$_2$X salts. \

However, above $\eta^{\ast}$, the AG formula is no more reliable and one should consider sizeable non-superconducting domains through which Cooper pairs tunnel between superconducting islands.
We have proposed a rather simple method to extract the key microscopic parameters from the experimental data to establish the relationship between our calculations and the experimental data. A good agreement has been found for various sources of disorder despite the simplifying assumptions.
The model could be improved by including the effect of a random distribution of the superconducting domains as observed recently in the high-$T_c$ superconductors La$_2$CuO$_4+y$ \cite{Bianconi} where a structural ordering of the oxygen interstitials has been reported. This order, which is highly inhomogeneous, is characterized by a fractal distribution which seems to enhance the superconducting transition temperature \cite{Bianconi}.
Such effect should be also relevant in (TMTSF)$_2$ClO$_4$ where previous study \cite{Pouget} has shown the presence of a large distribution of the ordered ClO$_4$ domain size in the relaxed sample. It has been found  that the width of this distribution shrinks as the cooling rate increases \cite{Pouget}.
We have also studied the role of superconducting fluctuations in layered superconductors with slab structure where transport measurements revealed non saturating upper critical field. We have ascribed this feature to the enhanced superconducting fluctuations to which transport probes are sensitive. Our results may explain the absence of any divergent behavior in the upper critical fields reported by thermodynamic measurements in organic superconductors.

\section{Acknowledgment}

We warmly thank D. J\'erome, C. Pasquier, J. G. Analytis, S. Blundell and P. Auban-Senzier S. Yonezawa for helpful and stimulating discussions.
This work was supported by the french-Tunisian CMCU project 10 G/1306.\newline
%

%

\end{document}